\documentclass[10pt,
			   aps,
			   pra,
			   twocolumn, 
			   nofootinbib, 
			   showkeys,           
			   superscriptaddress  
			   ]{revtex4-1}

\usepackage[english]{babel}
\usepackage[utf8]{inputenc}
\usepackage{amsmath, bm}
\usepackage[pdftex]{graphicx}
\usepackage{blindtext}
\usepackage{epsfig}
\usepackage{epstopdf}
\usepackage{scrextend}
\addtokomafont{labelinglabel}{\sffamily}
\usepackage[colorinlistoftodos]{todonotes}
\usepackage[left=1.5cm,right=1.5cm,
    top=1.5cm,bottom=1.5cm,bindingoffset=0cm]{geometry}

\newcommand{\bee}{\begin{equation}}
\newcommand{\ene}{\end{equation}}

\renewcommand{\phi}{\varphi}

\newcommand{\ve}[1]{{\bf #1}}

\newcommand{\om}{\omega}

\graphicspath{{pics/}}

\begin{document}
\title{Non-inverse dynamics of a quantum emitter  coupled to a fully anisotropic environment}

\date{\today}
\author{Danil Kornovan}
\affiliation{ITMO University}
\author{  Mihail Petrov}
\affiliation{ITMO University}
\author{ Ivan Iorsh}
\affiliation{ITMO University}

\begin{abstract}

 Anisotropic nanophotonic structures can couple the levels of a quantum emitter through the quantum interference effect. In this paper we study the coupling of quantum emitters excited states through the modes of a fully anisotropic structure: a structure for which all directions are physically nonequivalent. We consider an anisotropic metasurface as an illustrative example of such a structure. We point out a novel degree of freedom in controlling the temporal dynamics and spectral profiles of quantum emitters: namely, we show that a combination of the metasurface anisotropy and tilt of the emitter quantization axis with respect to the metasurface normal results in  nonsymmetric dynamics between the transitions of electrons from left-circular state to the right-circular states and the inverse process. Our findings give an additional mechanism for control over the light emission by quantum systems and, vice versa, can be utilized for probing  active transitions of quantum emitters. 

\end{abstract}

\maketitle
\section{Introduction}
The field of nanophotonics provides unique opportunities of controlling the  polarization state of light that governs the light-matter interaction. The non-zero optical spin moment of  electric field localized close to the structures interfaces allows for achieving  artificial chirality of light-matter interactions \cite{Lodahl2017}. Considered for the first time decades ago \cite{Gardiner1993,Carmichael1993}  the  chiral quantum optics   has acquired an experimental platform for observing chiral coupling of light with single quantum emitters  in photonic crystal waveguides \cite{Sollner2015,Coles2016}, nanofiber systems \cite{Corzo2016,Sorensen2016,Mitsch2014},  bottle microresonators \cite{Junge2013}, and planar grating systems \cite{Spitzer2018}. The nonsymmetric interaction of quantum emitters with the modes of nanophotonic structures makes  possible the effects of unidirectional quantum transport \cite{Kornovan2017a,Downing2019},  unusual optomechanical \cite{OudeWeernink2018, Kien2018}, and modifies the radiative properties of quantum ensembles \cite{Kornovan2017a,LeKien2017}.   Alternatively to atoms and quantum dots the semiconducting two-dimensional materials are a promising source of quantum chirality \cite{Wang2018} due to the circular optical transitions related to the spin-states of valley electrons. The important progress has been recently demonstrated in coupling of two-dimensional materials with plasmonic waveguides \cite{Su-Hyun2018} and  metasurfaces \cite{Chervy2018,Gong2018, Guddala2018,Hu2019}. The latter are naturally considered as    photonic counterparts to two-dimensional semiconductor materials. The metasurfaces (MS) have already demonstrated the unprecedented flexibility in the engineering of the polarization state of reflected and transmitted light~\cite{Glybovski2016} as well as localized surface waves \cite{Alu2015,Yermakov2015,Yermakov2016}, and  enhancement of the spontaneous emission rate of quantum sources \cite{Gomez-Diaz2015}. 

\begin{figure}[!h]
\includegraphics[width=0.99\linewidth]{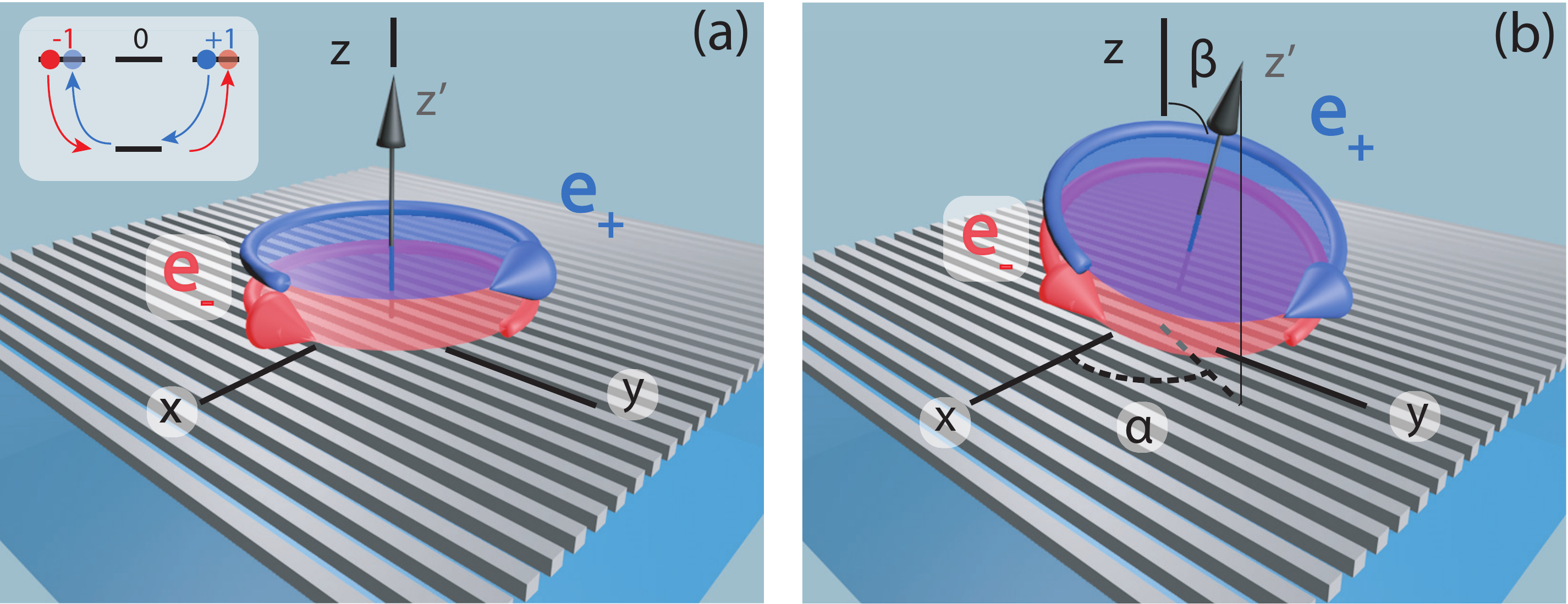}
\caption{ A general scheme of the set-up: a) a four-level atom with s$\to$p transition placed near an anisotropic metasurface. The rotating transition dipole moments of the atom are lie in the interface plane. The $x$ and $y$ axes are chosen in order to diagonalize the surface conductivity tensor $\sigma$. b) The same set-up but with the local quantization axis $z'$ rotated at  angles $\alpha, \beta$.}
\label{fig:scheme}
\end{figure}

The intrinsic anisotropy of nanophotonic interfaces results in the coupling of quantum transitions  \cite{Hughes2017,Jha2015} due to the quantum interference effect \cite{Agarwal2000}. Recently, it has been demonstrated that the coupling of orthogonal chiral states in two-dimensional materials with the  MS modes   leads to the formation of strongly coupled exciton-polariton states \cite{Chervy2018}, and coherence build up during the spontaneous transition was predicted \cite{Jha2018}. The possibility of an effective coupling of chiral transitions through the  MSs motivated us on studying the dynamics of transitions between two states with different total angular momentum projection, which become coupled due to the quantum interference enabled by the anisotropy of a metasurface (see Fig.~\ref{fig:scheme} (a)). We consider the  coupling of transitions with opposite helicities through the anisotropic MS, including the hyperbolic regimes. We predict that one can achieve non-inverse dynamics in transitions between the states with by tilting the local quantization axis of an emitter. The control over the orientation of the emitters quantization axis can be taken, for example, by applying a magnetic ~\cite{Oh1994} or electric field, which can be utilized in field sensing \cite{KoepsellPRA2017}. Moreover, orientation of a weak external magnetic field can control the spontaneous emission process from a multilevel atom into the modes of the structure \cite{KienPRA2008, KienPRA2017}. The orientation of transition dipole moments can  be also controlled by the strains induced in quantum dots, which has been demonstrated experimentally \cite{YuanNatCom2018}. The results proposed in this paper open a way for the reconstruction of the symmetry of the quantum emitter states based on its optical response, thus, realizing the optical tomography of quantum states with anisotropic metasurfaces.

\section{Metasurface induced quantum interference}
We start with a quantum emitter (QE) having optically allowed transition with three degenerate excited states $|e_{-1}\rangle, |e_{0}\rangle, |e_{+1}\rangle$ and a single ground state $|g\rangle$. The  corresponding transition dipole moments are denoted by $\mathbf{d_{-1}, d_{0}, d_{+1}}$ and their directions are given by the corresponding vectors: $\mathbf{e_{-1}} = +(\mathbf{e_x} - i \mathbf{e_y})/\sqrt{2}$, $\mathbf{e_0} = \mathbf{e_z}$, $\mathbf{e_{+1}} = -(\mathbf{e_x} + i \mathbf{e_y})/\sqrt{2}$, while the  amplitudes are assumed to be equal to $\mathbf{|d_{-1}|=|d_{0}|=|d_{+1}|}=d$. Note that at this stage the $\mathbf{d_{-1}}$, and $\mathbf{d_{+1}}$ are in the interface plane, while $\mathbf{d_0}$ is parallel to a normal of the structure. Having different angular momenta projections these states are orthogonal in the vacuum due to isotropy, but  placed in an anisotropic environment, these states may couple with each other via anisotropy induced quantum interference \cite{Agarwal2000}. An anisotropic metasurface is an example of such nanophotonic system,  breaking the isotropy in all three direction. The anisotropic response of the  metasurface can be well described \cite{Yermakov2015} within the effective conductivity tensor $\hat \sigma(\om)$. In  the coordinate system ({\it laboratory frame}) coinciding  with the main axes of the conductivity tensor, one gets: 
\begin{align}
\hat \sigma(\om)=\left(\begin{array}{cc}
\sigma_{xx}(\om)  & 0\\
0 & \sigma_{yy} (\om)\\
\end{array} \right),\nonumber 
\end{align}
where the diagonal entries are modelled with the Lorenzians $\sigma_{jj}(\omega) = A_{j} \dfrac{ic}{4\pi} \dfrac{\omega}{\omega^2 - \Omega_{j}^2 + i\gamma_{j}\omega}$ with $A_j$ being the normalization factor, $\Omega_j$ - resonance frequency, and $\gamma_j$ is the damping rate. Note that from now on we will use CGS units rather than SI.

The in-plane anisotropy of the metasurface allows for the coupling between the two transitions with opposite helicities [Fig.~\ref{fig:scheme} (a)]. Indeed,  the interaction between the states is described by the coupling constant $g_{-,+}$, which acquire non-vanishing values, namely:  $g_{-,+} = -4 \pi k_0^2 \ve{d_{-1}^*} \mathbf{G}(\mathbf{r_a}, \mathbf{r_a}, \omega_0) \mathbf{d_{+1}} / \hbar \sim G_{xx}(\ve{r_a}, \ve{r_a}, \om_0)-G_{yy}(\ve{r_a}, \ve{r_a}, \om_0)$, where $k_0$ is the wavenumber, and $\ve G(\ve{r_a}, \ve{r_a}, \omega_0)$ is the electromagnetic Green's tensor. The total Green's tensor consists of the vacuum and scattered contributions: $\mathbf{G} = \mathbf{G_0} + \mathbf{G_{sc}}$, it is known that for equal field and source points the real part of the vacuum contribution diverges. However, we will only take into account the imaginary part of it and consider the vacuum Lamb shift as being already included into the definition of emitter's transition frequency $\omega_0$.

The non-zero coupling between the states results in the redistribution of the quantum excitation if the system was initially pumped in any of the excited states. The temporal dynamics of  the system is governed by the evolution operator $U(t,0)$ \cite{ClaudeCohen-Tannoudji2004}, which gives the probability $P_{e_f,e_i}(t)=|U_{e_f e_i}(t,0)|^2$ for the atom to be in the excited state $|e_{f}\rangle$ at time $t$ given that it was in the state $|e_{i}\rangle$ initially. This allows us to implicitly reduce our attention to the subspace of the excited states only. 

The evolution operator can be expressed as:
\begin{eqnarray}
\langle e_{f}| \hat U(t,0) | e_{i} \rangle = \int_{C} \dfrac{dz}{2\pi i} e^{-i z t/\hbar} \langle e_{f}| \hat G(z) | e_{i} \rangle,
\label{evolmatrixel}
\end{eqnarray}
where $\hat G(z) = (z - \hat H)^{-1}$ is the resolvent operator of the full Hamiltonian $\hat H = \hat H_0 + \hat V$, consisting of the unperterbed $\hat H_0$ part and perturbation $\hat V$, and $|e_{i}\rangle, |e_{f}\rangle$ are the initial and final states, respectively. 

The unperturbed Hamiltonian consists of the field and atomic part $\hat H_0 = \hat H_A + \hat H_F$. In order to describe the field itself and its interaction with the atom for a very general case of a media with possible dispersion and absorption we employ the approach introduced in \cite{KnollArxiv2000}. In this case we can write:
\begin{eqnarray}
& \hat H_A = \sum\limits_{q=-1,0,+1} \hbar \omega_0 |e_q\rangle \langle e_q|, \nonumber\\
& \hat H_F = \int d\mathbf{r^{\prime}} \int\limits_0^{\infty} d\omega^{\prime} \hbar \omega^{\prime} \hat{\mathbf{f}}^\dagger (\mathbf{r^{\prime}},\omega^{\prime}) \hat{\mathbf{f}} (\mathbf{r^{\prime}},\omega^{\prime}), \nonumber\\
& \hat{V} = -\sum\limits_q \hat {\mathbf d}_{q} \hat {\mathbf E}(\mathbf{r_a}),	
\end{eqnarray}
where $\omega_0$ is the resonance frequency of the atomic transition, $\mathbf{\hat f^{\dagger}(\mathbf{r'}, \omega')}$ is the local field creation operator, $\mathbf{\hat E}(\mathbf{r_a})$ is the total electric field at the position of the atom $\mathbf{r_a}$. The electromagnetic field operator in this case reads as \\$\hat{\mathbf{E}}(\mathbf{r}) = i \; \sqrt[]{4 \hbar} \int d\mathbf{r^{\prime}} \int\limits_{0}^{\infty} d\omega^{\prime}\frac{{\omega^{\prime}}^2}{c^2} \sqrt{\varepsilon_I (\mathbf{r^{\prime}}, \omega^{\prime})}\mathbf{G}(\mathbf{r}, \mathbf{r^{\prime}}, \omega^{\prime}) \hat{\mathbf{f}}(\mathbf{r^{\prime}}, \omega^{\prime}) + h.c.$, where the bosonic field operators obey the commutation relation $\left[\hat{f}_i(\mathbf{r^{\prime}}, \omega^{\prime}), \hat{f}^{\dagger}_k(\mathbf{r}, \omega)\right] = \delta_{ik}\delta(\mathbf{r^{\prime}} - \mathbf{r})\delta(\omega^{\prime} - \omega)$, $\mathbf{G}(\mathbf{r}, \mathbf{r^\prime}, \omega^\prime)$ is the classical electromagnetic Green's function, and $\epsilon_{I}(\mathbf{r^\prime}, \omega^\prime)$ is the imaginary part of permittivity.

The resolvent operator $\hat G(z)$ projected onto the subspace of interest reads as:
\begin{eqnarray}
\hat P \hat G(z) \hat P = \hat P \dfrac{1}{z - \hat H_0 - \hat \Sigma(z)}\hat P,
\end{eqnarray}
here $\hat P$ is the projector onto the subspace, $\hat H_0$ is the unperturbed Hamiltonian and $\Sigma(z)$ is the level-shift operator or self-energy part. $\Sigma(z)$ here provides the energy shifts to the unperturbed eigenstates of $\hat H_0$ due to the interaction and has the form:
\begin{eqnarray}
\hat \Sigma(z)=\hat V + \hat V \hat G(z) \hat V \approx \hat V + \hat V \hat G_0(z=\hbar\omega_0) \hat V,
\end{eqnarray}
where the last equation implied the two approximations. The first one is near resonant case, which ignores possible dependence of $\hat \Sigma(z)$ on $z$, also called the flat spectrum approximation. The second one is that $\hat \Sigma(z)$ is calculated up to the second order in $\hat V$. 

The matrix elements of $\Sigma_{q',q}(\hbar \omega_0) = \langle e_{q'}| \hat \Sigma(\hbar \omega_0) | e_{q} \rangle$ represent the coupling of excited states through the modes of the field and can be found to be \cite{DungPRA2002_2}: $\Sigma_{q',q}(\hbar \omega_0)=-4\pi k_0^2\mathbf{d_{q'}^{*}}\mathbf{G}(\mathbf{r_a}, \mathbf{r_a},\omega_0)\mathbf{d_q}$, where $k_0=\omega_0/c$, $\mathbf{d_q}$ is the transition dipole moment.
\begin{figure}[htbp]
\includegraphics[width=0.9\linewidth]{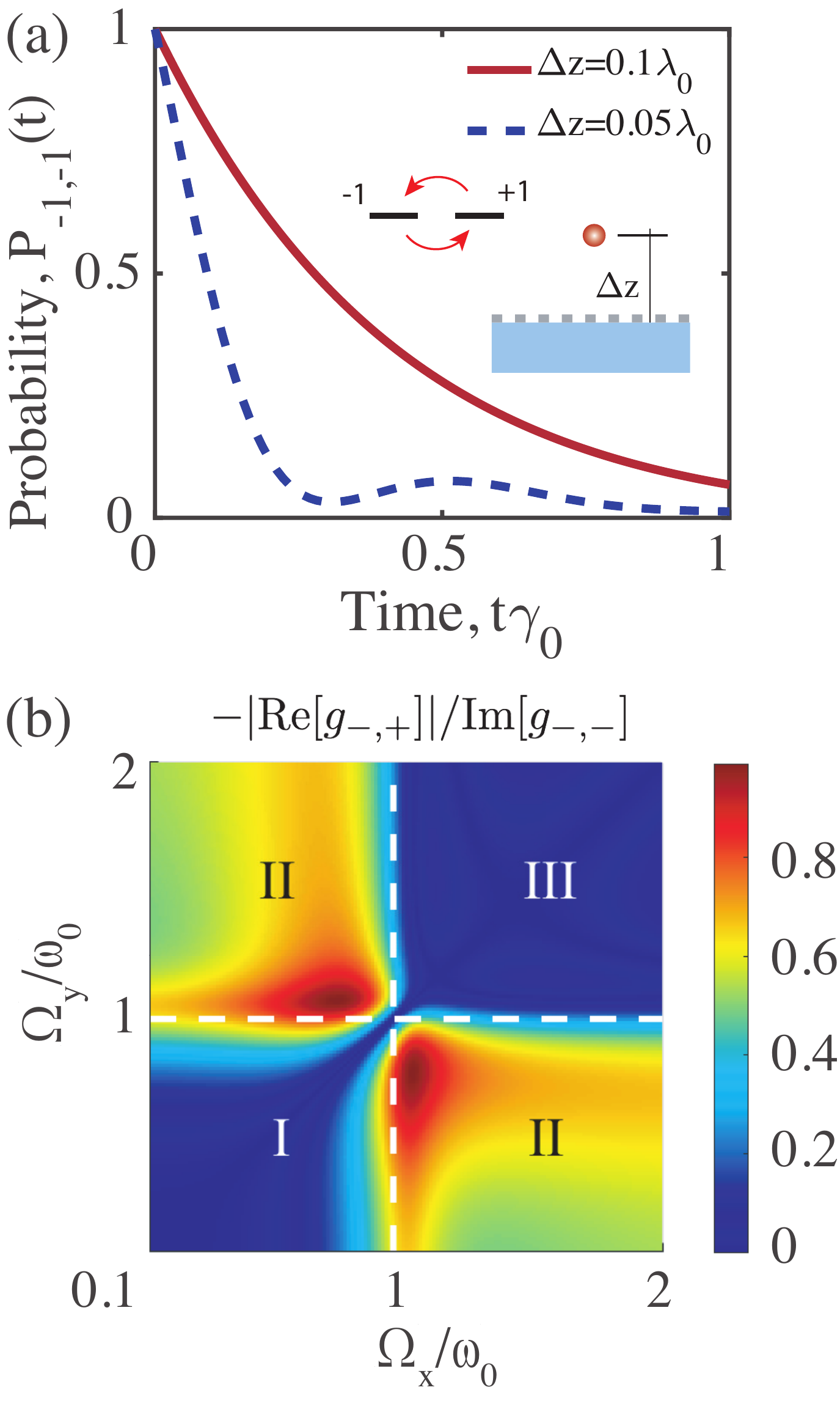}
\caption{a) Excited state probability $P_{-,-}(t)$ in the extreme anisotropic limit (see Eq. \eqref{CCSAMS}). The strong coupling regime  appears for an atom-metasurface distance $\Delta z = 0.05\lambda_0$. b) 2D map of the strong coupling parameter $-|\text{Re}[g_{-,+}]|/\text{Im}[g_{-,-}]$ as a function of $\Omega_x$, $\Omega_y$. The three specified regions correspond to: I) inductive ($\text{Im}[\sigma_{xx}], \text{Im}[\sigma_{yy}]>0$), II) hyperbolic ($\text{Im}[\sigma_{xx}]\cdot\text{Im}[\sigma_{yy}]<0$), and III) capacitive ($\text{Im}[\sigma_{xx}], \text{Im}[\sigma_{yy}]<0$) regimes. Other relevant parameters are: $\Delta z = 0.05\lambda_0$, $\gamma_x = \gamma_y = 0.1 \omega_0$, $\varepsilon_{subs} = 1$.}
\label{SAMS_TDyn}
\end{figure}
Once we construct the Green's tensor (see Appendix \ref{GTMS}) of a metasurface, we can compute the coupling elements $\Sigma_{q'q}(\hbar \omega_0)$ and solve for the dynamics of the atomic states population. In the set-up considered (Fig. \ref{SAMS_TDyn}, a) the states $|e_{-1}\rangle, |e_{+1}\rangle$ are mutually coupled while both being decoupled from $|e_{0}\rangle$ as a consequence of Green's tensor $\mathbf{G}(\mathbf{r_a}, \mathbf{r_a}, \omega)$ being diagonal.
Therefore, when studying the interaction between the states $|e_{+1}\rangle, |e_{-1}\rangle$, we can ignore $|e_0\rangle$ and immediately find the probability for the system to be in the state $|e_{-1}\rangle$ explicitly:
\begin{align}
	P_{{-1,-1}}(t,0) = \dfrac{1}{2} e^{2 \text{Im} [g_{-,-}] t} \left( \cos(2\text{Re}[g_{{-,+}}] t )\right. \nonumber\\+ \left.\cosh(2\text{Im}[g_{{-,+}}] t) \right),
	\label{probsymdyn}
\end{align}
 The  $g_{-,-}, g_{-,+}$  represent the diagonal and non-diagonal entries of ${\Sigma}(\hbar \omega_0)/\hbar$, respectively. The expression describing the dynamics of quantum states consists of two parts: a purely decaying term and  an oscillatory part. If one can achieve $\text{Re}[g_{-,+}] \ge -\text{Im}[g_{-,-}]$ then the oscillations will be underdamped, which corresponds to a \textit{strong coupling} of the  $|e_{+1}\rangle$ and $|e_{-1}\rangle$ states through the modes of the structure. 
 
It is illustrative to consider the case of the extreme anisotropy, when the $\sigma_{xx} \to 0 i$ and $\sigma_{yy} \to \infty i$, which corresponds to the ideal conductance in $y$-direction and isolation in $x$-direction. This gives us a very simple analytical result for the coupling constants:
\begin{align}
& \hbar g_{-,+} = \left( \dfrac{1 - i k \Delta z - 2 k^2 \Delta z^2}{4 \Delta z^3} \right)  |\mathbf{d}|^2 e^{i k 2 \Delta z}, \nonumber\\
& \hbar g_{-,-} = -\left( \dfrac{ i k \Delta z + 2 k^2 \Delta z^2}{4 \Delta z^3} \right)  |\mathbf{d}|^2 e^{i k 2 \Delta z}.
\label{CCSAMS}
\end{align}

In Fig. \ref{SAMS_TDyn}, (a) we plot the dynamics of the initially excited state for the case of the strong anisotropy in the absence (red solid) and the presence (blue dashed) of a strong coupling. However, for a more realistic case of finite loses and $\gamma_j \ne 0$, b) the strong coupling regime can \textit{almost} be achieved for the considered atom-surface distance $\Delta z = 0.05\lambda_0$  when at least one of the quantities $\Omega_x,\Omega_y$ is close to the atomic transition frequency $\omega_0$ only. One of the simplest ways to achieve the strong coupling is to consider smaller atom-surface distances $\Delta z$, however, at some point it might be necessary to consider also the Casimir-Polder interactions \cite{BuhmannDFII2012} with the modes of the nanostructure. 

The requirement for the strong coupling regime ($\text{Re}[g_{-,+}] \ge -\text{Im}[g_{-,-}]$) derived from Eq.~\eqref{probsymdyn} makes sense for the case of two interacting transitions. However, as we will show in the next section, it is still possible to achieve not only a measurable population transfer between the states but also non-equivalent dynamics for $|e_{-1}\rangle \to |e_{+1}\rangle$ and $|e_{+1}\rangle \to |e_{-1}\rangle$ processes. This can be done if the transition dipole moments are arbitrarily oriented and all three of them are coupled.

\section{Non-inverse dynamics of the four-level emitter}

The dynamics of the system is fully defined by the electromagnetic properties of the MS trough the coupling constants $g$ and is given by Eq.\eqref{probsymdyn}. The asymmetry in the quantum dynamics, i.e.when the dynamics of a transition from  $|e_{+1}\rangle \to |e_{-1}\rangle$ and $|e_{-1}\rangle \to |e_{+1}\rangle$ will be different, is also defined by the same coupling constants. Indeed, the dynamics of $|e_{-1}\rangle \to |e_{+1}\rangle$ transition can be obtained from the expression Eq.~\eqref{probsymdyn} by substituting $g_{-,+}\to g_{+,-} $. Thus, to get the asymmetry in the dynamics one should obtain the  difference in the  coupling constants.  From Green's function perspective the coupling constants read as: $g_{{}^{-,+}_{+,-}} \sim G_{yy} - G_{xx} \mp i\left( G_{xy} + G_{yx} \right)$. One can expect, that applying  a strong magnetic field along $z$-axis should break the time-reversal symmetry making the conductivity  tensor $\sigma$  to be  non-diagonal and antisymmetric~\cite{Gusynin2006}. This results in non-zero non-diagonal components of the Green's tensor $G_{xy}$ and $G_{yx}$ , however they have the opposite signs $G_{xy}=-G_{yx}$ which still makes the  coupling constants equal  $g_{-+}=g_{+-}$.


However, there is another, less direct way of breaking the symmetry between the $|e_{-1}\rangle, |e_{+1}\rangle$ states:  if one considers the atomic quantization axis to be tilted with respect to the laboratory axis $\mathbf{e_z}$ at an arbitrary angle as shown in Fig.1 (b).  Mathematically, this can be  described by considering the level-shift operator in the metasurface frame after the rotation of the quantization axis: 
$$\mathbf{\tilde \Sigma}(\hbar \omega_0) = \mathbf{S}^{\dagger}\ve M^{ \dagger}\mathbf{\Sigma}(\hbar \omega_0) \ve M  \mathbf{S}= \mathbf{T}^{\dagger}\mathbf{\Sigma}(\hbar \omega_0) \ve T.$$

\noindent Here $\mathbf{S}$ is the matrix of Cartesian components of the spherical tensors of rank $1$ as columns,  $\mathbf{M}$ is the  rotation matrix on Euler angles $\alpha, \beta, \gamma$ (active representation, right-hand rule, $z''-y'-z$ convention), and $\ve T= \ve M \ve S$ is the  composition of these transformations.   



All of the information about the dynamics of the system is contained in the eigenvalues and eigenstates of $\ve{ \tilde \Sigma}(\hbar \omega_0)$.
   Note that the rotation of the quantization axis does not change the eigenvalues, but alters the eigenstates of the system, and the evolution matrix can be expressed as:
\begin{eqnarray}
 U_{q' q}(t,0) = \sum\limits_{j=x,y,z} C_{j}^{(q',q)} e^{-i g_{j} t}; \: \: \:   q,q'=\{-1,0,+1\},
 \label{UDyn}
\end{eqnarray}
 where $g_{j}$ are the complex eigenvalues of $\mathbf{\Sigma}$, and $C_{j}^{(q',q)} = \left(T^{-1}\right)_{q',j} T_{j,q}$. We should note that since our nanostructure is the planar conducting interface, the physical meaning of eigenstates of the system are the three linear dipole moments aligned along highly symmetric directions of the environment, therefore, $\hbar g_{j} = - 4 \pi k_0^2 \ve{d_j^\dagger} \ve{G}(\ve{r_a}, \ve{r_a}, \omega) \ve{d_j}$ are the self-couplings of three linear dipole moments oriented along $x$, $y$, and $z$. 
 
 The coefficients $C^{(q',q)}_j$ for the case of the excitation transfer from $|e_{+1}\rangle \to |e_{-1}\rangle$ have the following explicit form:
\begin{eqnarray}
	& C^{(-1,+1)}_{x} = -\frac{e^{-2i\gamma}}{2} \left( \cos(\alpha) \cos(\beta) - i \sin(\alpha) \right)^2 , \nonumber\\
	& C^{(-1,+1)}_{y} = \frac{e^{-2i\gamma}}{2} \left( \cos(\alpha) - i \cos(\beta) \sin(\alpha) \right)^2 , \nonumber\\
	& C^{(-1,+1)}_{z} = - \frac{e^{-2i\gamma}}{2} \sin^2(\beta).
	\label{CCoef}
\end{eqnarray} 

\begin{figure}[htbp]
	\begin{center}
		\includegraphics[width=0.9\linewidth]{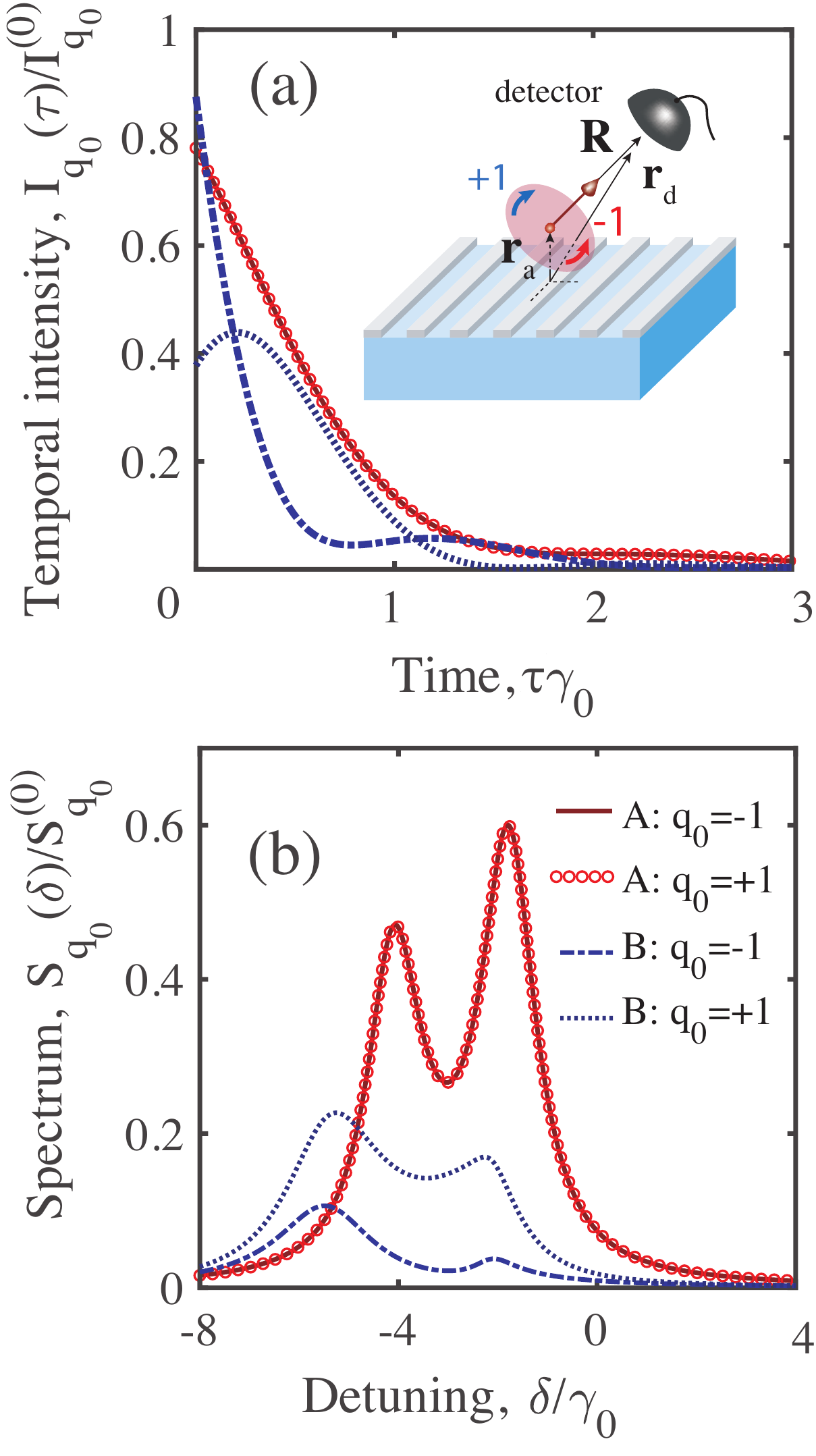}
	\end{center}
	\caption{a) Local field intensity registered at the detector position $\mathbf{r_d}$ versus time $\tau = t - R/c$ measured in the units of a free space emission rate $\gamma_0$. Zero time corresponds to a moment when the emitted light reaches the detector. The atom is initially in the  $|e_{q_0}\rangle$ state. The two cases are studied: case A - isotropic ($\Omega_x=\Omega_y=1.5k_0$) and B - anisotropic ($\Omega_x=1.5k_0, \Omega_y=1.1k_0$) metasurface. There are also two initial conditions considered with 4 cases in total: isotropic case - $A: q_0 = -1$ and $A: q_0 = +1$ (solid dark red line and bright red circles); anisotropic case - $B: q_0 = -1$ and $B: q_0 = +1$ (blue dash-dotted and dotted line, respectively). The parameters are: $\Omega_{x}=0.6 k_0$, $\Omega_{y}=1.0 k_0$, $\gamma_{x} = \gamma_{y} = 0.1 k_0$, $\epsilon_{subs} = 1.0$, $\Delta z = 0.05 \lambda_0$, $\mathbf{r_a} = (0, 0, \Delta z)$, $\mathbf{r_d} = R (\cos(\alpha) \sin(\beta), \sin(\alpha) \sin(\beta), \cos(\beta))$, $R = 100 \lambda_0$, $\alpha = \pi/4$, $\beta = \pi/4$. Normalization factor $I^{(0)}_{q_0}$ is the intensity detected at moment $t = R/c$ in the absence of a metasurface. b) The total emitted light spectra. The relevant parameters and the cases considered are the same as for Fig. \ref{IntTD} (a), $S_{q_0}^{(0)}$ is the resonant value of the total emitted light spectrum for an atom in the vacuum. The comments on how we choose parameters for A and B are given in Appendix \ref{MDP}. }
	\label{IntTD}
\end{figure}

 First of all, one can notice that since $\gamma$ enters as an overall phase for all of the coefficients $C_{j}^{(q',q)}$ then it does not affect the population dynamics $P_{q',q}(t,0) = |U_{q',q}(t,0)|^2$. It means that the last rotation around the new  $\mathbf{e_z'}$ axis at angle $\gamma$ is redundant and without loss of generality we can set $\gamma = 0$. From the unitarity  of $\ve T(\alpha, \beta, \gamma)$ it follows that $C^{(+,-)}_{j} = (C^{(-,+)}_{j})^*$. This fact immediately gives us a straightforward result that for arbitrary values of  $\alpha, \beta$ one can achieve non-inverse dynamics between the two transition   as the coefficients are not equal anymore. 
Indeed, the asymmetry in the excitation transport between the states manifests itself in the following way:
\begin{align}
&	P_{-,+}(t) - P_{+,-}(t) =\nonumber \\
& f(\alpha, \beta) \sum\limits_{(k,l)}^{(x,y),(y,z),(z,x)} \:  e^{(g_k''+g_l'')t} \sin\left( \left( g_k' - g_l' \right)t \right),
\label{DynDisc}
\end{align}
where $g_{j} = g_{j}' + ig_{j}''$ , and $f(\alpha, \beta) = \frac{1}{8} \sin(2\alpha) \sin(2\beta) \sin(\beta)$. Equation~\eqref{DynDisc} is the central result of the paper, and it has several important  consequences. First, one can see that the difference  vanishes in two cases: i) when $f(\alpha, \beta) = 0$ or, equivalently, when either $\alpha$ or $\beta$ are integer multiples of ${\pi}/{2}$; ii) when any two directions are equivalent, so that $g_{k} = g_{l}$. Thus, to obtain the non-inverse dynamics of the excitation we need to both have the atomic system quantization axis to be tilted at an arbitrary angle relative to the metasurface and have full anisotropy of the environment. 

We want to stress that this effect can be observed for the process involving any pair of states with $q,q' = \{ -1,0,+1\}$, not only the processes coupling the $|e_{-1}\rangle$ and $|e_{+1}\rangle$. It is also important to mention that from the form of transport asymmetry \eqref{DynDisc} one can formulate the explicit way to describe the difference of $P_{-1,+1}(t)$ and $P_{+1,-1}(t)$. Indeed, we can calculate the actual probailities given by $P_{q',q}(t)=|U_{q',q}(t,0)|^2$:
	
\begin{multline}
P_{q',q}(t) = \sum_{k=x,y,z} C_{k,k}^{(q',q)} e^{2 g_j'' t} + \\
\sum\limits_{(k,l)}^{(x,y),(y,z),(z,x)} 2 |C_{k,l}^{(q',q)} | \cos\left[ ( g_k' - g_l')t - \phi_{k,l}^{(q',q)} \right] e^{(g_k'' + g_l'')t},
\end{multline}
where $g_j$ are, as in eq. \eqref{UDyn}, the eigenvalues of $\Sigma_{q',q}(\hbar \omega_0)/\hbar$ (self-couplings of linear dipoles along $x, y, z$), $C_{k,l}^{(q',q)} = C_k^{(q',q)} (C_l^{(q',q)})^*$, and $\phi_{k,l}^{(q',q)} = \arg \left( C_{k,l}^{(q',q)} \right)$. Note that the second sum is responsible for the interference of contributions from different eigenstates. From the property $C_{k,l}^{(q',q)} = \left( C_{k,l}^{(q,q')} \right)^*$ immediately follows that $\phi_{k,l}^{(q',q)} = - \phi_{k,l}^{(q,q')}$, which means that the difference in population transfer probabilities $P_{q',q}(t)$ and $P_{q,q'}(t)$ manifests itself as the \textit{phase delay} in the interference part of the dynamics.

In this section we described the physical origin of the effect under study in terms of internal degrees of freedom of the emitter - transition probabilities $P_{q',q}(t)$. In the next section we proceed by considering how the observable quantities like detected light intensity or emitted spectrum are affected, which might be of special interest if one keeps in mind a possible experimental verification.



\section{The effect on the measurable observables}

\subsection{Far-field intensity dynamics}
The temporal dynamics can  be detected by measuring the far-field radiation generated by the atom. Basing on the results presented in Ref.~  \cite{DungPRA2000, DungPRA2002}, we will obtain the detected light intensity, the temporal profile of which is given by \cite{DungPRA2002}:
\begin{eqnarray}
I_{q_0}(t) = \Bigg| \sum\limits_{q'} 4 k_0^2 \int\limits_0^t dt' C_{q',q_0} (t') \int\limits_{0}^{\infty} d \omega \text{Im} \left[ \mathbf{G}(\mathbf{r_d}, \mathbf{r_a}, \omega) \right] \mathbf{d_{q'}} \nonumber\\
e^{-i (\omega - \omega_0) (t - t')} \Bigg|^2, \quad \quad
\label{IntDef}
\end{eqnarray}
with $\mathbf{r_d}, \mathbf{r_a}$ being the position of the detector and atom, respectively, $C_{q',q_0}(t)$ is the probability amplitude that state $q'$ is excited at time $t$, while initially the system was in $q_0$ state.

If we want to find the intensity detected in the far-field zone, we need to replace the full Green's tensor with its far-field part $\mathbf{G} (\mathbf{r_d}, \mathbf{r_a}, \omega) \to \mathbf{G^{FF}} (\mathbf{r_d}, \mathbf{r_a}, \omega)$. According to a superposition principle \cite{Chew1999, Tai1994} this far-field GF can be written as a sum of the free-space and the scattered part: $\mathbf{G^{FF}} (\mathbf{r_d}, \mathbf{r_a}, \omega) = \mathbf{G^{FF, 0}} (\mathbf{r_d}, \mathbf{r_a}, \omega) + \mathbf{G^{FF, sc}} (\mathbf{r_d}, \mathbf{r_a}, \omega)$ or:

\begin{eqnarray}
\mathbf{G^{FF}} (\mathbf{r_d}, \mathbf{r_a}, \omega) = \mathbf{f^{0}} (\mathbf{r_d}, \mathbf{r_a}, \omega) e^{i k R_{-}}  + \mathbf{f^{sc}} (\mathbf{r_d}, \mathbf{r_a}, \omega) e^{i k R_{+}}, \nonumber\\
{ }
\end{eqnarray}
where $R_{\pm} = \sqrt{ (x_d - x_a)^2 + (y_d - y_a)^2 + (z_d \pm z_a)^2 }$. The phases in the exponent differ as these two contributions are created by two dipoles: one is located at the position $\mathbf{r_a}$ and the other one is its mirror image located at $\begin{pmatrix}
x_a, y_a, -z_a
\end{pmatrix}$. However, if we put a dipole very close to a surface so that $z_0/\lambda_0 \ll 1$ then we can ignore this discrepancy and set $R_{-} = R_{+} = R$, obtaining that $\mathbf{G^{FF}} (\mathbf{r_d}, \mathbf{r_a}, \omega) \approx \mathbf{f} e^{i k R}$. Now using $\text{Im}\left[ \mathbf{f}  (\mathbf{r_d}, \mathbf{r_a}, \omega) e^{i k R} \right] = \text{Re} \left[ \mathbf{f}  (\mathbf{r_d}, \mathbf{r_a}, \omega) \right] \sin(k R) + \text{Im}\left[ \mathbf{f}  (\mathbf{r_d}, \mathbf{r_a}, \omega) \right] \cos(k R)$ and making the expansion near the resonance frequency $k(\omega) \approx k(\omega_0) + k'(\omega_0) (\omega - \omega_0)$, we can proceed by taking the frequency integral in eq. \eqref{IntDef}. The $\mathbf{f}(\mathbf{r_d}, \mathbf{r_a}, \omega)$ function in our case can be regarded as slowly varying function of frequency and put in front of the $\omega$ integral taken at resonance frequency $\omega_0$. Then we can perform the $\omega$ integral and arrive at the following result:
\begin{eqnarray}
I_{q_0}(t) \approx \left| \dfrac{4 \pi k_0^2 }{i} \sum\limits_{q'} C_{q',q_0}(t - R/c) \mathbf{G^{FF}} (\mathbf{r_d}, \mathbf{r_a}, \omega_0) \mathbf{d_{q'}} \right|^2.
\label{IntDef2}
\end{eqnarray}

One should note that this form naturally expresses the total amplitude as a sum of contributions from 3 dipole moments associated with each active transition $q' = (-1, 0, +1)$ multiplied by the probability amplitude of the corresponding excited state at the retarded time $\tau = t - R/c$. We can rewrite eq. \eqref{IntDef2} by making use of $\mathbf{d_q}$ definition and eq. \eqref{UDyn} in the form:
\begin{eqnarray}
& I_{q_0}(\tau) \approx | 4 \pi k_0^2 \mathbf{d} |^2 \bigg| \sum\limits_j \mathbf{f_j} e^{-i g_j \tau} \bigg|^2,
\label{IntDynEigS}
\end{eqnarray}
here $\tau = (t - R/c)$, $c$ is the speed of light, $\mathbf{f_j} = \mathbf{G^{FF}_{:, j}} (\mathbf{r_d}, \mathbf{r_a}, \omega_0
) \left( \mathbf{M S}_{j, q_0} \right)$ is related to the field generated at the detector's position $\mathbf{r_d}$, $\ve r_a$ is the position of a QE, and  $\mathbf{G^{FF}_{:,j}}(\mathbf{r_d}, \mathbf{r_a}, \omega_0)$ is the j$^{\text{th}}$ column of the far-field classical Green's tensor of the system. It is clear that the temporal dynamics described by Eq.~\eqref{UDyn} is directly mapped onto this quantity.

%

In order to observe the manifestation of the effect it is convenient to compare the intensity dynamics for the two initial conditions (an atom being in $| e_{-1} \rangle$ and $| e_{+1} \rangle$ initially) in the case of isotropic and anisotropic metasurfaces. The results are presented in Fig. \ref{IntTD} (a). Notice that the position of the detector is rotated with respect to the axis origin in the same way as the atomic local quantization $z$ axis is rotated: it is simply $\mathbf{r_d}|| \mathbf{M} \hat{z}$. This keeps the number of degrees of freedom constant as the orientations of atomic quantization axis and detectors position relative to an atom are now related and described only by $(\alpha, \beta)$. One can notice in Fig.\ref{IntTD} (a) that for the set-up considered for the anisotropic metasurface there is a difference in temporal dynamics of the detected field intensity $I_{q_0}(\tau)$ for initially excited states with opposite helicities $A: q_0=-1$ (blue dashed-dotted) and $A: q_0=+1$ (blue dotted). For the isotropic case the difference between the $B: q_0=-1$ (solid dark red line) and $B: q_0=+1$ (bright red circles) intensity profiles vanishes, as expected.

It is also important to mention that it might look natural to consider a hyperbolic regime for a metasurface when it comes to studying the light-matter interactions as the surface plasmon-polariton (SPP) modes are prominent in this case. Despite the fact the SPPs might have a very strong local field (leading to the increase of $\text{Re}[g_{j}]$), they also carry the energy away from the system due to strongly enhanced spontaneous emission (and, therefore, high $-\text{Im}[g_j]$).  One can conclude that for the problem considered in our work, the {\it near field modes} which are forbidden to propagate in any direction are of interest, but not the propagating modes.

\subsection{Far-field emitted light spectrum}

In the previous section we studied how the temporal intensity profile is affected by the described phenomenon. In order to get the insights into the spectral manifestation of the aforementioned asymmetry, we calculate the far-field spectrum of the initially excited atom. According to \cite{DungPRA2002} one can find the emitted light spectrum in the Markov approximation:
\begin{eqnarray}
& S_{q_0}(\omega) = \left| \sum\limits_{q'} \int\limits_0^\infty dt' C_{q',q_0}(t') e^{i(\omega - \omega_{q'}) t'} \mathbf{F^{q'}}(\mathbf{r_d}, \mathbf{r_a}) \right|^2,
\label{SpecDungDef}
\end{eqnarray}
where $\omega_{q'}$ is the transition frequency $|g\rangle \to |e_{q'}\rangle$, and $C_{q',q_0}(t')$ is the excite state $|e_{q'}\rangle$ probability amplitude (with $q_0$ being the initial state) given by \eqref{UDyn}, and $\mathbf{F^{q'}}(\mathbf{r_d}, \mathbf{r_a})$ is :

\begin{multline}
\mathbf{F^{q'}}(\mathbf{r_d}, \mathbf{r_a}) = \\
4 \dfrac{\omega_{q'}^2}{c^2} \int' d\omega' \text{Im}[\mathbf{G}(\mathbf{r_d}, \mathbf{r_a}, \omega')] \mathbf{d_{q'}} \zeta (\omega_{q'} - \omega') = \mathbf{\Gamma}(\mathbf{r_d}, \mathbf{r_a}) \mathbf{d_{q'}},
\label{FieldSpec}
\end{multline}
with  $\zeta(x) = i \text{P} \frac{1}{x} + \pi \delta(x) $. 

The unrotated transition dipole moments are given by $\mathbf{d_{q'}} = |\mathbf{d}|\mathbf{S}_{:, q'}$, where $\mathbf{S} = \left[ \mathbf{e_{-1}}, \mathbf{e_{0}}, \mathbf{e_{+1}} \right]$ - matrix, where each column is a spherical tensor in Cartesian coordinates. We need to rotate each vector written in Cartesian coordinates on Euler angles using the matrix $\mathbf{M}(\alpha, \beta, \gamma)$, so the rotated vectors are $\mathbf{M} \mathbf{S_{:,q'}}$.

Also, unlike in \cite{DungPRA2002} to make it more coherent with our work we consider that the unperturbed Hamiltonian  does not take into account the Lamb shift for each excited state and we account for it in $\mathbf{\Sigma}_{rot}$. Formally, it means that in the Green's function argument $\omega_0$ has to be replaced by the corrected atomic transition frequency with the Lamb shift included. We need to note that this can not lead to any significant changes as the corresponding corrections are on the order of $\sim \gamma_0$, while we consider that $\omega_0 \gg \gamma_0$ and the Green's tensor in our problem varies significantly on the frequency range on the order of $\omega_0$.

We also take only the $\pi \delta(x)$ part of $\zeta(x)$ in \eqref{FieldSpec} for the sake of simplicity. Even though the principal value part can be significant, it will not affect the result qualitatively.

Now, doing the integral over $t'$ in \eqref{SpecDungDef} we arrive at the form of \eqref{FullSpec}:

\begin{widetext}
	\begin{eqnarray}
	& S_{q_0} (\omega) = \left| \sum\limits_{j}  \dfrac{i \sum_{q'} \mathbf{F_{q'} (\mathbf{r_d}, \mathbf{r_a})} C_j^{(q', q_0)} }{(\delta - g_{j})} \right|^2 = \left| \sum\limits_{j}  \dfrac{i \sum_{q'} \mathbf{\Gamma}(\mathbf{r_d}, \mathbf{r_a}) \mathbf{d_{q'}}  C_j^{(q', q_0)} }{(\delta - g_{j})} \right|^2 = \nonumber\\
	& \left| \sum\limits_{j}  \dfrac{i \sum_{q'} |\mathbf{d}| \mathbf{\Gamma}(\mathbf{r_d}, \mathbf{r_a}) \mathbf{MS}_{:, q'} \left[(\mathbf{MS})^{-1}\right]_{q', j} (\mathbf{MS})_{j, q_0} }{(\delta - g_{j})} \right|^2 = \left| \sum\limits_{j}  \dfrac{i |\mathbf{d}| \mathbf{\Gamma}_{:, j} (\mathbf{r_d}, \mathbf{r_a}) (\mathbf{MS})_{j, q_0} }{(\delta - g_{j})} \right|^2 = \left| \sum\limits_j \dfrac{\bm{f}_{j}^{q_0}(\mathbf{r_d}, \mathbf{r_a})}{ (\delta - g_j) } \right|^2,
	\end{eqnarray}
\end{widetext}

where we used the fact that $C_{j}^{(q', q_0)}$ is nothing but $\left[ (\mathbf{MS})^{-1}\right]_{q', j} (\mathbf{MS})_{j, q_0}$. Another way of representing the total spectrum can be obtained if we do the following:

\begin{widetext}
	\begin{eqnarray}
	& \sum\limits_{j=1}^{N} \sum\limits_{i=1}^{N} \dfrac{ ( \bm{f_i^{q_0}})^{\dagger} \bm{f_j^{q_0}} }{(\delta - g_j)(\delta - g_i^*)} = \sum\limits_{j=1}^{N} \sum\limits_{i=1}^{N} \dfrac{ ( \bm{f_i^{q_0}})^{\dagger} \bm{f_j^{q_0}} }{g_j - g_i^*} \left[ \dfrac{1}{\delta - g_j} - \dfrac{1}{\delta - g_i^*} \right] = \nonumber\\
	& \sum\limits_{j=1}^{N} \sum\limits_{i=1}^{N} 2 \: \text{Re} \left[ \dfrac{ ( \bm{f_i^{q_0}})^{\dagger} \bm{f_j^{q_0}} }{ g_j - g_i^* } \dfrac{1}{\delta - g_j} \right] = \sum\limits_{j=1}^{N} \sum\limits_{i=1}^{N} 2 \: \text{Re} \left[ \dfrac{ ( \bm{f_i^{q_0}})^{\dagger} \bm{f_j^{q_0}} (\delta - g_j^*)}{g_j - g_i^*} \right] \dfrac{1}{|\delta - g_j|^2},
	\end{eqnarray}
\end{widetext}
where in transition from the 1st to the 2nd line the interchange $i \leftrightarrow j$ in the second term was made. 

We can define the following two quantities:
\begin{eqnarray}
& \xi_j = + 2 \text{Re}\left[ \sum\limits_i \dfrac{ (\bm{ f_i^{q_0}})^\dagger \bm{f_j^{q_0}} }{g_j - g_i^*} \right], \eta_j = - 2 \text{Im} \left[  \sum\limits_i \dfrac{ (\bm{ f_i^{q_0}})^\dagger \bm{f_j^{q_0}} }{g_j - g_i^*}  \right], \nonumber\\
{ }
\end{eqnarray}
and, finally, obtain the following:
\begin{eqnarray}
 S_{q_0}(\delta) &\approx& \left| \sum\limits_j \dfrac{\bm{f}_{j}^{q_0}(\mathbf{r_d}, \mathbf{r_a})}{ (\delta - g_j) } \right|^2  =\nonumber \\
& &  \sum\limits_{j} \dfrac{ \left( \xi_j (\delta - g_j') + \eta_j g_j'' \right)} {(\delta - g_j')^2 + g_j''^2},
\label{FullSpec}
\end{eqnarray}
The last part of Eq.~\eqref{FullSpec} simply shows that the spectrum for 3 eigenstates can be decomposed into 3 symmetric Lorentzian lines $\sim \eta_j$ and 3 antisymmetric parts $\sim \xi_j$, and depending upon their values the lineshape can vary significantly. 

Here we also want to note that it is easy to calculate the emitted spectrum of a certain polarization. Formally, this can be done by replacing the vector $\bm{f}_{j}^{q_0}(\mathbf{r_d}, \mathbf{r_a})$ with a scalar $f_{j}^{q_d, q_0}(\mathbf{r_d}, \mathbf{r_a}) = \mathbf{e_{q_{d}}^{\dagger}} \cdot \bm{f}_{j}^{q_0}(\mathbf{r_d}, \mathbf{r_a})$, where $q_d$ is polarization to which the detector is sensitive, and $\mathbf{e_{q_{d}}^{\dagger}}$ is the corresponding normalized polarization vector. In this case the total emitted light spectrum is simply $S_{q_0}(\delta) = \sum_{q_d} S_{q_d,q_0}(\delta)$ as a result of the completeness relation $\sum_{q_d} \mathbf{e_{q_d}} \otimes \mathbf{e_{q_d}^{\dagger}} = \mathbf{1}$.

As expected, the total emitted light spectra also differ for two initially excited states of opposite helicities in case of an anisotropic structure, and with the tilted atomic quantization axis, see Fig. \ref{IntTD}, (b). It should be stressed that tilting the quantization axis is not the only way to observe the difference between $S_{-1}(\delta)$ and $S_{+1}(\delta)$.  Namely, the introduction of the substrate with $\varepsilon_{subs}\neq 1$ leads to a similar result. However, this happens not due to different populations of atomic energy levels, but rather due to the mixing of the fields emitted by different eigenstates. The details can be found in Appendix \ref{SubsC}.

The described effect opens the route towards the optical tomography of the internal state of the quantum emitters. Namely, placing an isolated emitter or an array of them in the vicinity of the structure would allow the reconstruction of the symmetry axes of the nanosized object by the scattering spectra or transient radiation dynamics. The effect under study is also of importance for spectroscopy and has to be taken into account. 

\section{Conclusion}
We have shown that the combined effect of the metasurface anisotropy and the tilt of quantum emitter quantization axis leads to an observable difference both in temporal dynamics and also spectral properties of the emitter initially pumped into states of opposite helicities. This is a somewhat counterintuitive result since it states that optical activity can emerge due to the anisotropy of the system and it originates through the quantum interference of the multiple decay channels of the emitter.

The results presented here are applicable not only for metasurfaces but for any structure that is fully anisotropic, for example, planar cavities with in-plane anisotropy or an ensemble of ultracold atoms trapped near the optical nanofiber \cite{MitschNatCom2014}. Moreover, in the case of cavity it should be enhanced by the order of the quality factor while the field localization is usually smaller for the case of cavities than for the metasurfaces or waveguide structures. These findings not only open new avenues towards the engineering of the quantum optical states at the nanoscale, but can be readily used for the relatively simple optical tomography of the nanoobjects.


\section{Acknowledgement}
We thank for fruitful discussions A.A. Bogdanov, L.E. Golub, M.M. Glazov, O.E. Yermakov.

This work was supported by the Russian Foundation for Basic Research (pojects \# 18-32-00691, \# 17-02-01234).
D. K. and M.P. acknowledge the support from Basis Foundation.

\bibliographystyle{unsrt}
\bibliography{main.bib}

\appendix

\section{Green's tensor of a metasurface} 
\label{GTMS} 

The total Green's tensor according to a superposition principle \cite{Chew1999, Tai1994} of the problem can be expanded into the following sum:
\begin{eqnarray}
	\mathbf{G^{ij}}(\mathbf{r}, \mathbf{r'}, \omega) = \mathbf{G_0}(\mathbf{r}, \mathbf{r'}, \omega) + \mathbf{G_{sc}^{ij}}(\mathbf{r}, \mathbf{r'}, \omega),
\end{eqnarray}
where $\mathbf{G_0}(\mathbf{r}, \mathbf{r'}, \omega)$ is the free space Green's tensor, and $\mathbf{G_{sc}}(\mathbf{r}, \mathbf{r'}, \omega)$ is the scattered part, which contains all the information about the modes of the structure. The superscipts $^{ij}$ describe the position of the field and the source points with respect to the interface of the structure. We label the upper halfspace as $1$ and the lower one as $2$. We are especially interested in constructing the $\mathbf{G^{11}}(\mathbf{r}, \mathbf{r'}, \omega)$ tensor, and we also want to find $\mathbf{G^{21}}(\mathbf{r}, \mathbf{r'}, \omega) = \mathbf{G^{21}_{sc}}(\mathbf{r}, \mathbf{r'}, \omega)$ to satisfy the boundary conditions on the interface.

In order to find the scattered part we want to use the approach described in \cite{Lakhtakia1992} and begin by introducing the following vector functions, corresponding to TE/TM modes:
\begin{eqnarray}
\mathbf{t_{j,\pm}} = \dfrac{1}{\kappa} \begin{pmatrix}
-\kappa_y \\
+\kappa_x \\
0
\end{pmatrix}, \quad \mathbf{p_{j, \pm}} = \dfrac{1}{k_j} \begin{pmatrix}
\mp k_{j,z} \kappa_x/\kappa \\
\mp k_{j,z} \kappa_y/\kappa \\
\kappa
\end{pmatrix},
\end{eqnarray} 
here $\kappa = \sqrt{\kappa_x^2 + \kappa_y^2}; \quad k_{j,z} = \sqrt{k_j^2 - \kappa^2}$. The first subscript (in $\mathbf{t/p}$ functions $_j$) labels the media, while the $_{\pm}$ defines the propagation direction along the $z$ axis. 

The expansions for both free and the scattered parts have the form:
\begin{widetext}
\begin{eqnarray}
&\mathbf{G_0}(\mathbf{r}, \mathbf{r'}, \omega) = -\mathbf{e_z}\mathbf{e_z}\delta(\mathbf{R}) + \dfrac{i}{8 \pi^2} \int \int d\kappa_x d\kappa_y \dfrac{1}{k_{1z}} \left[ \mathbf{t_{1\pm}} \mathbf{t_{1\pm}} + \mathbf{p_{1\pm}} \mathbf{p_{1\pm}}\right] \exp(i\mathbf{k_{1\pm}\mathbf{R}}), \nonumber\\
&\mathbf{G_{sc}^{11}}(\mathbf{r}, \mathbf{r'}, \omega) = \dfrac{i}{8\pi^2}\int \int \dfrac{d\kappa_x d\kappa_y}{k_{1,z}} \bigg[ R^{11}_{tt} \mathbf{t_{1, +}} \mathbf{t_{1, -}} + R^{11}_{tp} \mathbf{t_{1, +}} \mathbf{p_{1, -}} + R^{11}_{pt} \mathbf{p_{1, +}} \mathbf{t_{1, -}} + R^{11}_{pp} \mathbf{p_{1, +}} \mathbf{p_{1, -}}\bigg] \exp(i\mathbf{k_{1,+}}\mathbf{r} - i\mathbf{k_{1,-}}\mathbf{r'}), \nonumber\\
&\mathbf{G_{sc}^{21}}(\mathbf{r}, \mathbf{r'}, \omega) = \dfrac{i}{8\pi^2}\int \int \dfrac{d\kappa_x d\kappa_y}{k_{1,z}} \bigg[ R^{21}_{tt} \mathbf{t_{2, -}} \mathbf{t_{1, -}} + R^{21}_{tp} \mathbf{t_{2, -}} \mathbf{p_{1, -}} + R^{21}_{pt} \mathbf{p_{2, -}} \mathbf{t_{1, -}} + R^{21}_{pp} \mathbf{p_{2, -}} \mathbf{p_{1, -}}\bigg] \exp(i\mathbf{k_{2,-}}\mathbf{r} - i\mathbf{k_{1,-}}\mathbf{r'}), \nonumber\\
\end{eqnarray}
{ }
\end{widetext}
here $\mathbf{R} = \mathbf{r} - \mathbf{r'}$. In $\mathbf{G_0}(\mathbf{r}, \mathbf{r'}, \omega)$ the upper (lower) signs in the field vector functions are for the case $z>z'$ $(z<z')$. Here we also introduced the Fresnel coefficients $R^{ij}_{kl}$ accounting for the scatterring of the mode "l" into the mode "k". Note that since our structure is, in general, anisotropic in the $xy$ plane, there are modes of a hybrid nature which can be identified by the cross terms involving products of $\mathbf{t_{j,\pm}}$ and $\mathbf{p_{j,\pm}}$. 

The coefficients $R^{ij}_{kl}$ can be found by satisfying the boundary conditions for both electric and magnetic fields:
\begin{eqnarray}
\begin{cases}
\mathbf{e_{z}} \times \left( \mathbf{E_{1}} - \mathbf{E_{2}}\right) = 0, \\
\mathbf{e_{z}} \times \left( \mathbf{H_{1}} - \mathbf{H_{2}} \right) = \dfrac{4\pi}{c} \mathbf{\sigma} \mathbf{E_{1,2}},
\end{cases}
\label{boundarycond}
\end{eqnarray}
where $\mathbf{\sigma}$ is a surface conductivity tensor. 
The first condition on the electric field allows to relate different Fresnel coefficients to each other in a rather simple form:
\begin{eqnarray}
&1 + R^{11}_{tt} = R^{21}_{tt} , \nonumber\\
&R^{11}_{pt} \dfrac{k_{1,z}}{k_1} = -R^{21}_{pt} \dfrac{k_{2,z}}{k_2} , \nonumber\\
&R^{11}_{tp} = R^{21}_{tp} , \nonumber\\
& - 1 + R^{11}_{pp} = - R^{21}_{pp} \dfrac{k_{2,z} k_1}{k_2 k_{1,z}}.
\end{eqnarray}

By using this along with the second line of \eqref{boundarycond} we can find the rest of the coefficients.

The optical properties of such a metasurface can be characterized by the tensor of the effective surface conductivity $\sigma$, which can be chosen to be diagonal in some reference frame. To describe the optical properties of a metasurface we use the effective conductivity described by \cite{YermakovPRB2015}:
\begin{equation}
\sigma = \begin{pmatrix}
\sigma_{xx} & 0 \\
0 & \sigma_{yy}
\end{pmatrix}, \quad \sigma_{jj} = A_{j} \dfrac{ic}{4\pi} \dfrac{\omega}{\omega^2 - \Omega_{j}^2 + i\gamma_{j}\omega},
\end{equation}

where $A_{j}$ is the normalization constant, $\Omega_{j}$ is the resonance frequency, $\gamma_{j}$ is the bandwidth. 

In the case of the absent substrate ($\epsilon_1 = \epsilon_2$) and strong anisotropy ($\sigma_{yy} \to i\infty,  \sigma_{xx} \to i0$), we can obtain the $\mathbf{G^{11}_{sc}(\mathbf{r}, \mathbf{r}, \omega)}$ analytically:
\begin{eqnarray}
& G_{xx}^{sc,11}(\mathbf{r}, \mathbf{r}, \omega) = \dfrac{1}{32 \pi k^2 \Delta z^3} e^{i k 2 \Delta z} , \nonumber\\
& G_{yy}^{sc,11}(\mathbf{r}, \mathbf{r}, \omega) = \dfrac{-1 + 2 i k \Delta z + 4 k^2 \Delta z^2}{32 \pi k^2 \Delta z^3} e^{i k 2 \Delta z} , \nonumber\\
& G^{sc,11}_{zz}(\mathbf{r}, \mathbf{r}, \omega) =   \dfrac{1 - ik\Delta z}{16 \pi  \Delta z^3 k^2} e^{i k 2 \Delta z} .
\end{eqnarray}

\section{A measure of the discrepancy between the intensity and spectral profiles}
\label{MDP} 

As we are interested in both intensity $I_{q_0}(\tau)$ and spectrum $S_{q_0}(\delta)$ for two initial conditions ($q_0 = -1,+1$), it might be good to study how the difference between these two cases depends upon the metasurface parameters. For this we need to fix the orientation of the local quantization z-axis (angles $\alpha, \beta$). Now we introduce the following two quantities:

\begin{eqnarray}
& \tilde{I}_{-1,+1} = \dfrac{ \int\limits_{0}^{\infty} \left| I_{-1}(\tau) - I_{+1}(\tau) \right| d\tau}{\int\limits_{0}^{\infty} \left( I_{-1}(\tau) + I_{+1}(\tau) \right) d\tau}, \label{Itilde} \\
& \tilde{S}_{-1,+1} = \dfrac{ \int\limits_{-\infty}^{\infty} \left| S_{-1}(\delta) - S_{+1}(\delta) \right| d\delta}{\int\limits_{-\infty}^{\infty} \left( S_{-1}(\delta) + S_{+1}(\delta) \right) d\delta}.
\label{STilde}
\end{eqnarray}

\begin{figure}[t]
	\begin{center}
		\includegraphics[width=0.235\textwidth]{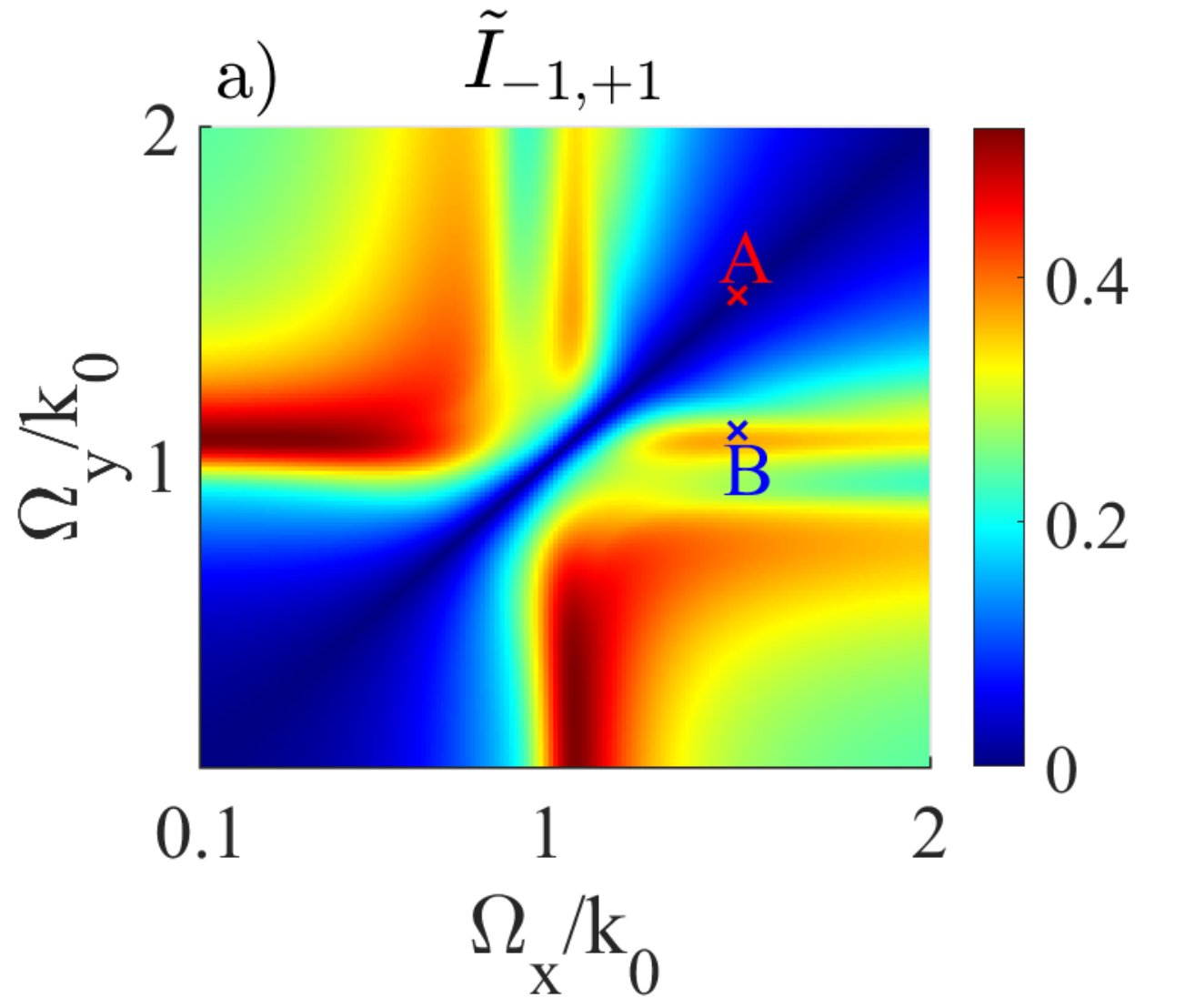}
		\includegraphics[width=0.235\textwidth]{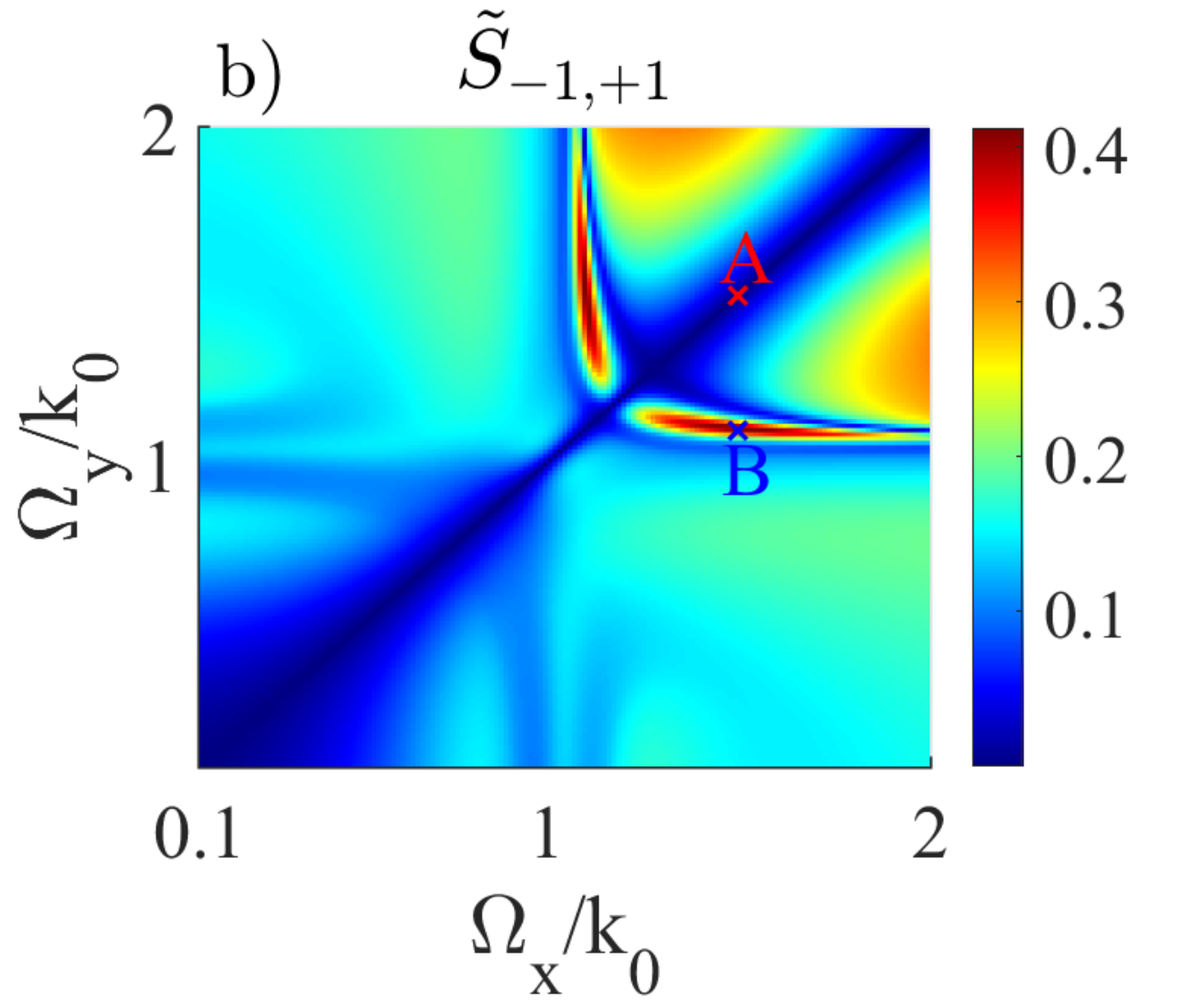}
	\end{center}
	\caption{$\tilde{I}_{-1,+1}$ and $\tilde{S}_{-1,+1}$ parameters defined by \eqref{Itilde}, \eqref{STilde} versus metasurface resonance frequencies $\Omega_x$, $\Omega_y$. All other parameters are the same as in Fig. \ref{IntTD}. The two specific points correspond to an isotropic (A: $\Omega_x = \Omega_y = 1.5k_0$) and anisotropic ($\Omega_x = 1.5k_0$, $\Omega_y = 1.1k_0$) cases.}
	\label{ISTilde}
\end{figure}

Clearly, these two quantities are always between $0$ and $1$ and can be used to measure how much the two graphs are similar or different. Therefore, we can plot the map of \eqref{Itilde}, \eqref{STilde} versus resonance frequencies $\Omega_x, \Omega_y$ presented in Fig. \ref{ISTilde} a, b. Note that for $\Omega_x=\Omega_y$ the metasurface is isotropic and both $\tilde{I}_{-1,+1}$, $\tilde{S}_{-1,+1}$ are equal to zero. Even though the local maxima of $\tilde{I}_{-1,+1}$ and $\tilde{S}_{-1,+1}$ do not overlap, there is a region where both of these quantities exceed the value of $\sim 0.2-0.3$, which is sufficient in order to observe the discrepancy (see Fig. \ref{ISTilde} a and b). 

\section{Emitted light intensity and spectrum in the case of a substrate.}
\label{SubsC} 

\begin{figure}[t]
	\begin{center}
		\includegraphics[width=0.238\textwidth]{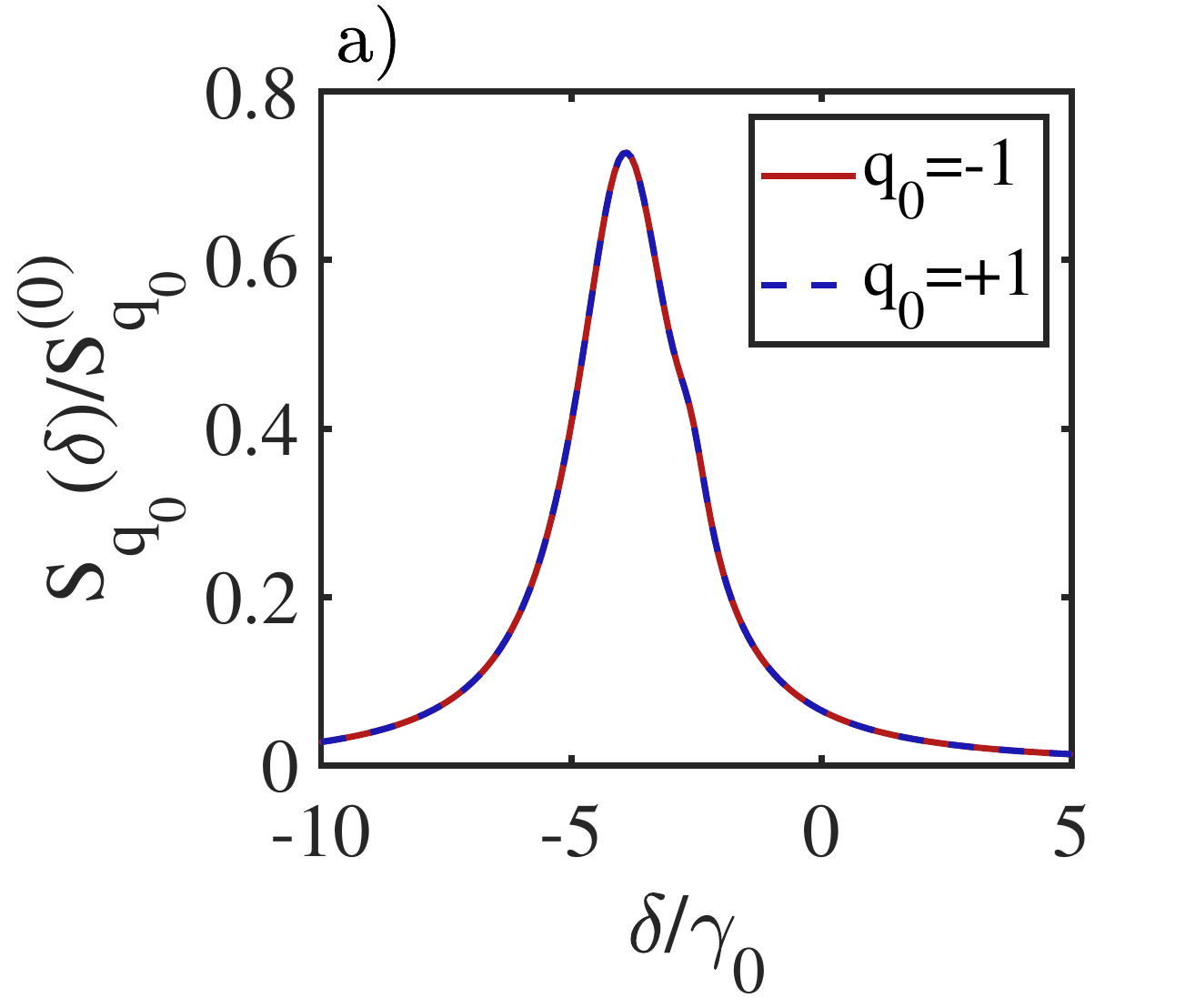}
		\includegraphics[width=0.238\textwidth]{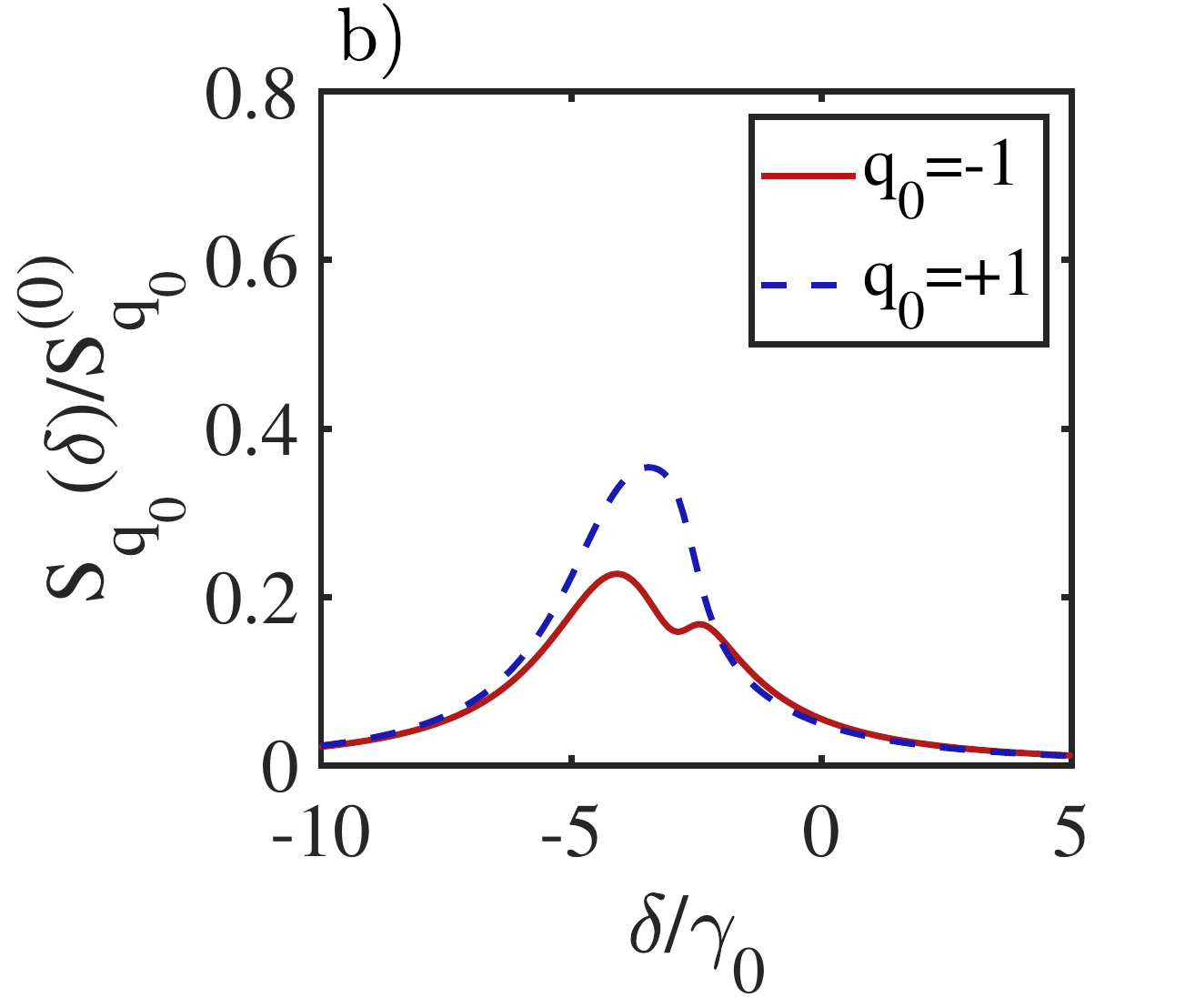}
	\end{center}
	\caption{The total emitted light spectra for two initial conditions: $q_0 = -1$ (solid red lines) and $q_0 = +1$ (dashed blue lines) in case of an absent substrate $\varepsilon_{subs} = 1$ (a) and with substrate $\varepsilon_{subs} = 2.2$ (b) for in-plane situation ($\alpha = \beta = 0$). The other relevant parameters are the same as for Fig. \ref{IntTD}, $S_{q_0}^{(0)}$ is the resonant value of the total emitted light spectrum for an atom in the vacuum.}
	\label{TotSpecInPlane}
\end{figure}

One major difference between observing the probabilities of some processes $P_{e_f,e_i}(t)$ and looking at either detected intensity $I_{q_0}(t)$ or spectrum $S_{q_0}(\delta)$  is that in two latter cases the position of the detector with respect to the atom and nanostructure is involved. One can consider the case when $\mathbf{d_{-1}}, \mathbf{d_{+1}}$ rotate in the interface plane ($\alpha = \beta = 0$) and put detector right above the atom into the far field so that it has the position $\mathbf{r_d} = \left( 0, 0, R\right)$. Note that in this scenario the eigenstate with the $z$-oriented associated dipole moment does not contribute to the result as it does not have the far-field term. If there is no substrate $\epsilon_{subs} = 1$ then the corresponding Green's tensor $\mathbf{G^{FF}(\mathbf{r_d}, \mathbf{r_a}, \omega_0)}$ is diagonal and the two relevant contributions in Eq. \eqref{IntDynEigS}, \eqref{FullSpec} from $x$ and $y$ dipole moments do not interfere with each other. However, if one introduces the substrate $\epsilon_{subs} \ne 1$ then there are non-zero components of the Green's tensor $G^{FF}_{xy}(\mathbf{r_d}, \mathbf{r_a}, \omega_0) = G^{FF}_{yx}(\mathbf{r_d}, \mathbf{r_a}, \omega_0) \ne 0$, which leads to the mixing of the fields generated by $x$ and $y$ dipoles resulting in the observable difference in $S_{-1}(\delta), S_{+1}(\delta)$. However, one should not confuse this with the effect described in the maintext as in this scenario the transition probabilities will be equal: $P_{-,+}(t) = P_{+,-}(t)$. 

Indeed, as can be seen from Fig. \ref{TotSpecInPlane}, the presence of the substrate breaks the symmetry between the $q_0 = -1$ and $q_0 = +1$ cases leading to an observable difference in the emitted light spectrum for instance.

\end{document}